\documentclass[aps,prl,twocolumn]{revtex4}

\usepackage{graphicx}
\usepackage{dcolumn}
\usepackage{bm}
\usepackage{amsmath}
\usepackage{amssymb}
\usepackage{amsfonts}
\usepackage{amsthm}
\usepackage{graphicx}
\usepackage{hyperref}
\usepackage{color}
\usepackage{array} 
\usepackage{multirow}
\usepackage{subfigure}
\usepackage{accents}

\newcommand{\braket}[2]{\mbox{$ \langle #1 | #2 \rangle $}}

\newcommand{\ket}[1]{\mbox{$ | #1 \rangle $}}

\newcommand*{\dt}[1]{\accentset{\mbox{\large\bfseries .}}{#1}}

\begin{document}

\title{Observation of quantum interference as a function of Berry's phase
\\ in a complex Hadamard optical network}

\author{Anthony Laing}
\email{anthony.laing@bristol.ac.uk}
\author{Thomas Lawson}
\author{Enrique Mart\'{i}n L\'{o}pez}
\author{Jeremy L. O'Brien}
\affiliation{Centre for Quantum Photonics, H. H. Wills Physics Laboratory \& Department of Electrical and Electronic Engineering, University of Bristol, BS8 1UB, United Kingdom}

\begin{abstract}
Emerging models of quantum computation driven by multi-photon quantum interference, while not universal, may offer an exponential advantage over classical computers for certain problems. Implementing these circuits via geometric phase gates could mitigate requirements for error correction to achieve fault tolerance while retaining their relative physical simplicity. We report an experiment in which a geometric phase is embedded in an optical network with no closed-loops, enabling quantum interference between two photons as a function of the phase.
\end{abstract}

\maketitle
When a quantum mechanical system evolves under some Hamiltonian, the probability amplitudes associated with indistinguishable events can accumulate dynamical and geometric phases \cite{Berry} and interfere constructively or destructively.  In Hong Ou Mandel interference \cite{ho-prl-59-2044}, when two photons meet at the input ports of a beamsplitter, the event that both photons are transmitted is indistinguishable from the event that both photons are reflected, but the associated probability amplitudes have opposite phases so interfere destructively: the probability to detect one photon at each output port is zero.  This quintessentially \emph{quantum} photonic interference generates the non-classical correlations in multi-photon quantum walks \cite{br-prl-102-253904, pe-sci-329-1500} and the computational complexity of many-photon interference in large optical networks \cite{va-tcs-8-189, sc-arxiv-2004, aaronsonQIP2010}. These emerging models of quantum computation are unlikely to be universal, but may be exponentially more powerful than classical computers for certain problems.  Crucially, since the basic models do not require initial entanglement, conditional gates, or feed-forward operations, large scale examples will be substantially less challenging to physically construct than a universal quantum computer. Achieving fault tolerance in these schemes without sacrificing their relative physical simplicity to unwieldy error correction is a key goal. 

Geometric phases and, more generally, non abelian holonomies have been proposed as a method to implement fault-tolerant gates for \emph{universal} quantum computation \cite{holonomic_proposal,pa-pra-61-010305,geometric_NMR, dcz01}, since they are robust against perturbations to which the important global geometric properties are invariant \cite{ca-prl-90-160402,de-prl-91-090404,ca-prl-92-020402,so-pra-70-042316,fu-prl-94-020503,fi-prl-102-030404}.  As described by Berry \cite{Berry}, a phase is accrued by a vector in an instantaneous eigenstate of a Hamiltonian undergoing cyclic adiabatic evolution.  Anticipated in an earlier classical result \cite{pancharatnam1956} and verified experimentally \cite{to-prl-57-937, ch-ol-13-562, le-sci-318-1889} the geometric phase has undergone important generalisations, including the non-abelian \cite{wi-prl-52-2111} and non-adiabatic cases \cite{ah-prl-58-1593}, the non-cyclic \cite{jo-pra-38-1590, wu-prb-38-11907, we-prl-64-1318} and non-unitary \cite{sa-prl-60-2339} cases, and the case where the endpoints of the evolution are orthogonal \cite{ma-prl-85-3067, la-prl-72-1004, ha-prl-87-070401}.  The geometric phase has been observed at the single photon level and in the context of non-locality \cite{kw-prl-66-588, st-pra-56-3129, er-prl-94-050401}, while a bi-photon wavepacket in a superposition of modes in a closed interferometer exhibits the predicted increase in sensitivity to a geometric phase \cite{kl-pla-140-19, br-pra-52-2551, ko-pra-83-063808} that is observed for a dynamical phase \cite{ra-prl-65-1348}.  To date, however, all observations of optical geometric phases involve self-interference of a single photon or classical light interference.

In light of the computational attributes of quantum interference between different photons and the desire to achieve fault tolerance in physically feasible computational models driven by this effect, demonstrating exquisite control over photonic quantum interference via an intrinsically robust geometric phase gate is a key step.  Such an experimental connection in the context of these models is somewhat analogous to the implementation of a holonomic two-qubit gate in the established circuit model of universal quantum computation.  Furthermore, to directly observe the influence of the geometric phase on interference between photons, any measurement statistics should not be obfuscated by other phase dependent phenomena.  In particular, single-photon interference, which has already been demonstrated to be predictably receptive to the geometric phase, should ideally be independent from the geometric phase in such an observation.

Here, we establish an experimental functional relationship connecting a variable geometric phase (vGP) to sinusoidal quantum interference between individual photons of a pair.  The vGP is imparted inside a four mode optical network that contains no closed loops, such that no single-photon interference can take place.  Applied to only one photon of the pair in one of the modes, the vGP arises through a traversal of the polarisation-sphere comprising a closed cycle and an open path; the end points of the total traversal are mutually orthogonal.  The other three modes traverse lengths on polarisation-sphere equal to that of the vGP mode, but these include a partial, or total, path retracing such that a fixed GP, or no GP, is finally imparted.  Polarisation vectors evolve via parallel transport, ensuring zero dynamical phase here, while the unknown dynamical phase contributions from small physical length mismatches are fixed.  We observe high visibility quantum interference fringes, and find an approximate flat line response for one-photon inputs, confirming the absence of single-photon interference.

Complex Hadamard matrices \cite{ta-osid-13-133} relate the computational basis of a discrete Hilbert space to some \emph{mutually unbiased} basis \cite{du-ijqi-8-535}.  The four mode complex Hadamard unitary, $H_{4}$, transforms quantum states according to
\begin{equation*}
H_{4}= \frac{1}{2}
 \left( \begin{array}{cccc}
1 & 1 & 1 & 1 \\
1 & e^{i \theta} & -1 & -e^{i \theta} \\
1 & -1 & 1 & -1\\
1 & -e^{i \theta}  & -1 & e^{i \theta}
\end{array} \right)
\label{eqQCH}
\end{equation*}
so that a single particle prepared in a well defined position corresponding to an element of the computational basis, when acted upon by a device described by $H_{4}$, emerges with maximal uncertainty in its position.  A large ensemble of similarly prepared one-particle input states will be found after the device with approximately one quarter of their total number at each of the four detectors (in the case of no losses).  Modulation of the phase $\theta$ in the $H_{4}$ device has no consequence for the maximal uncertainty in position of the single particle, and no consequence for the detection statistics of the ensemble.

\begin{figure}[t]
\subfigure{\includegraphics[trim=0 0 0 0, clip,width=.77\columnwidth]{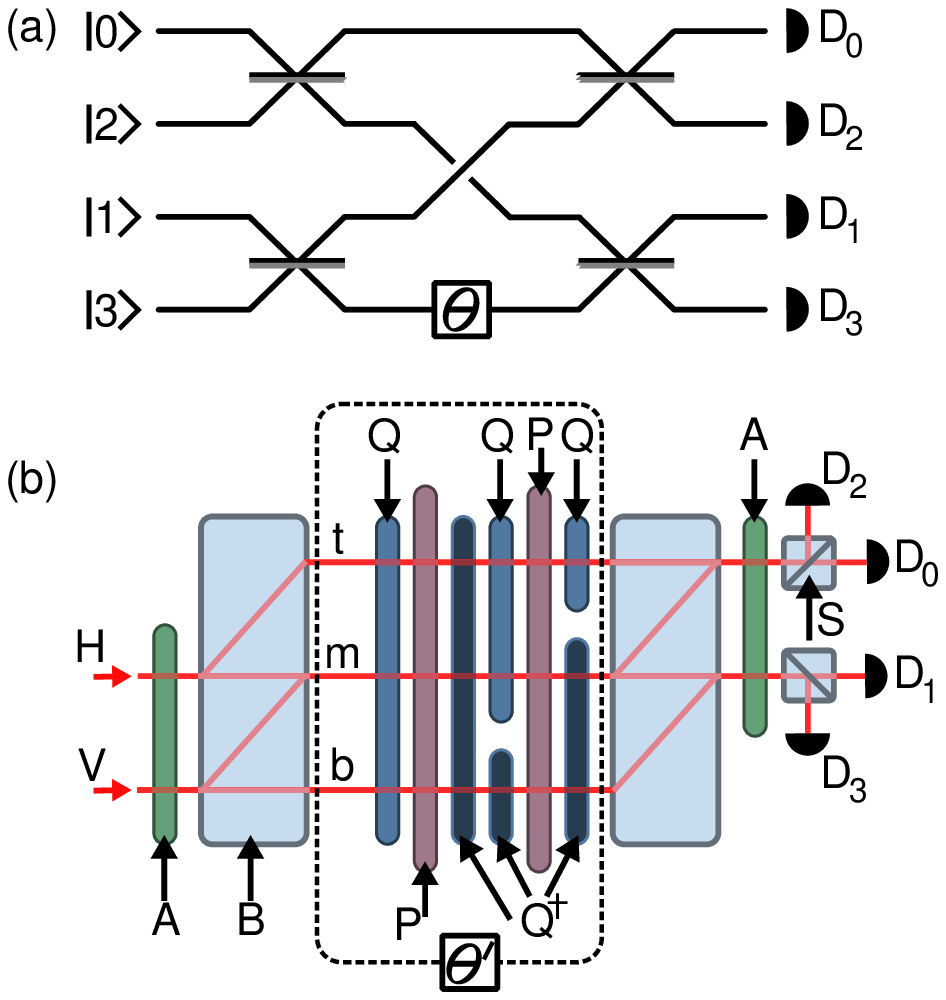}}\\
\vspace{-4 pt}
\subfigure{\includegraphics[trim=10 16 -10 20, clip,width=.63\columnwidth]{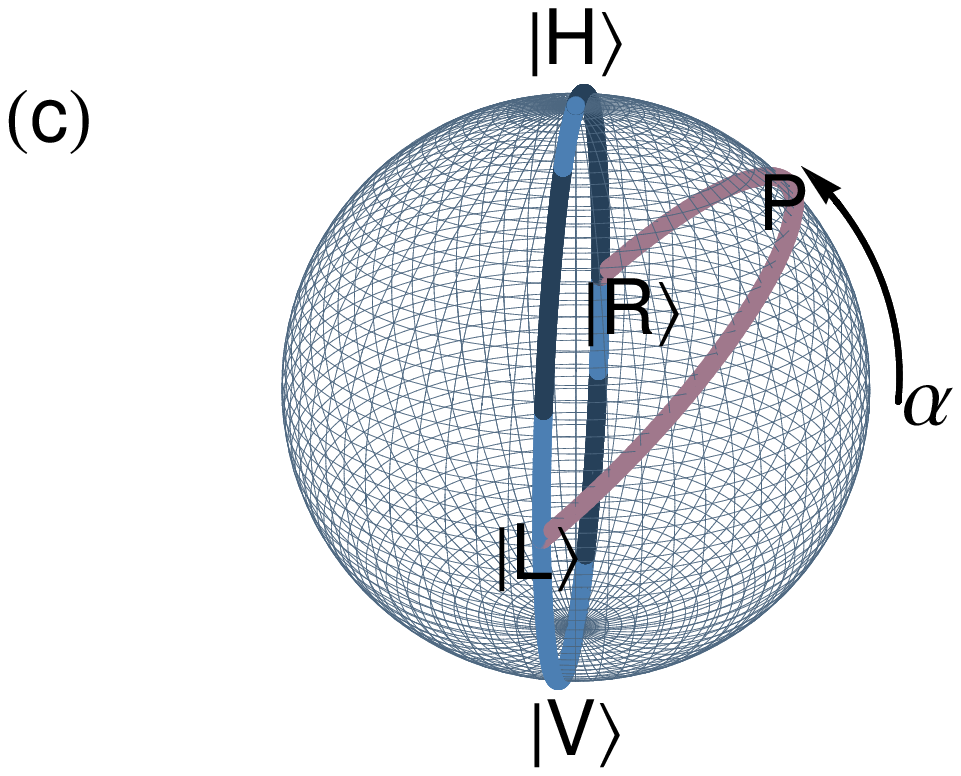}}
\vspace{-6 pt}
\caption{Complex Hadamard network and vGP. (a) Schematic of the optical network, with phase shift $\theta$. (b) Experimental construction based on beam displacers, with waveplates implementing a geometric phase $\theta^{\prime}$. (c) Sphere showing stages of polarisation vector travel in the $\theta^{\prime}$ section.}
\label{fgCirc}
\vspace{-10 pt}
\end{figure}

The optical network shown schematically in Fig.~\ref{fgCirc}(a) consists of four one-half reflectivity beamsplitters, a swap of the two middle modes, a phase shift $\theta$ on the lowest mode, and four detectors ($D_{i}$); it is equivalent to $H_{4}$ with the labelling of input and output ports indicated.  The network contains no closed-loop interferometers and a photon injected into any input port emerges in an equal superposition of the four output ports, regardless of the phase setting.  Similarly, measurement in the computational basis of one-photon ensembles cannot reveal any information about $\theta$, and these statistics should ideally give a flat line response to modulation of this phase.

The situation is dramatically different for two-photon input states.  Simultaneous injection into the complex Hadamard network of a photon in mode $\ket{0}$ and a photon in mode $\ket{1}$, leads to a state that experiences photonic quantum interference, producing correlations between pairs of detectors as a function of $\theta$.  The conditional probabilities for coincidental detection of photons are summarised in Eq.~\eqref{eqProbs}.  Given a detection at $D_{i}$, the probability $\Pr( j \mid i )$ for a detection $D_{j}$ is 
\begin{align}
\Pr(i \pm 1 \!\!\! \pmod{4} \mid i=\{0,2\}) &= (1\pm \cos{\theta})/2\nonumber
\\[4pt]
\Pr(i + 2 \mid i=\{0,1\}) &= 0.
\label{eqProbs}
\end{align}
Implementing $\theta$ geometrically creates a system in which quantum correlations from two-photon inputs are a function of a geometric phase, whereas its effects are unobservable in statistics from one-photon inputs, which would not be the case in an interferometer such as the Mach Zehnder.

\begin{figure*}[t]
\centering
\subfigure{\includegraphics[trim=205 10 201 0, clip,width=0.96\textwidth]{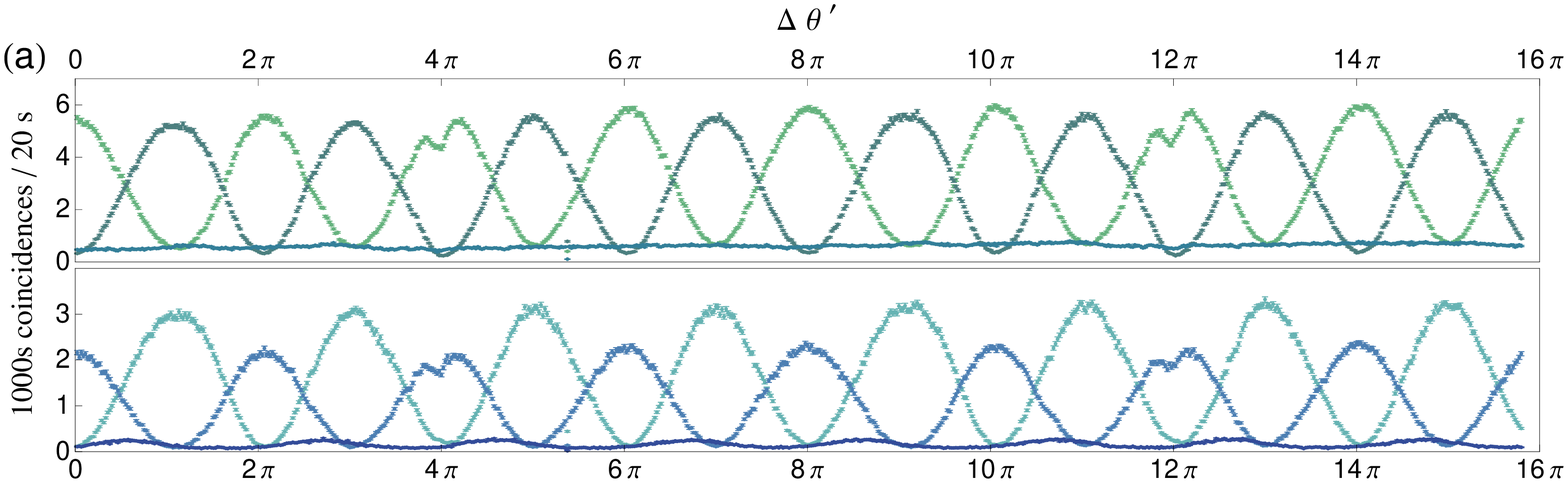}}\\
\vspace{-10 pt}
\subfigure{\includegraphics[trim=12 10 20 20, clip,width=0.32\textwidth]{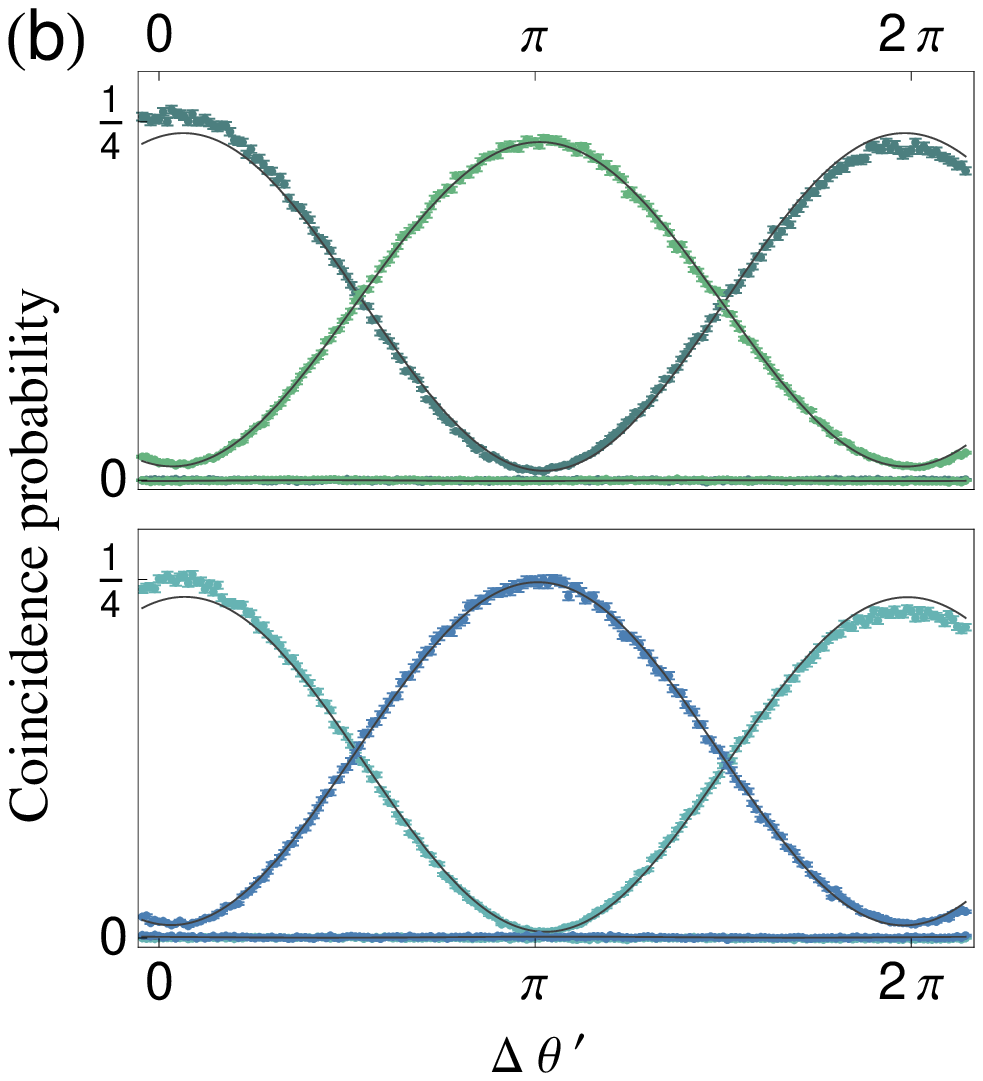}}
\subfigure{\includegraphics[trim=20 36 160 150, clip,width=0.64\textwidth]{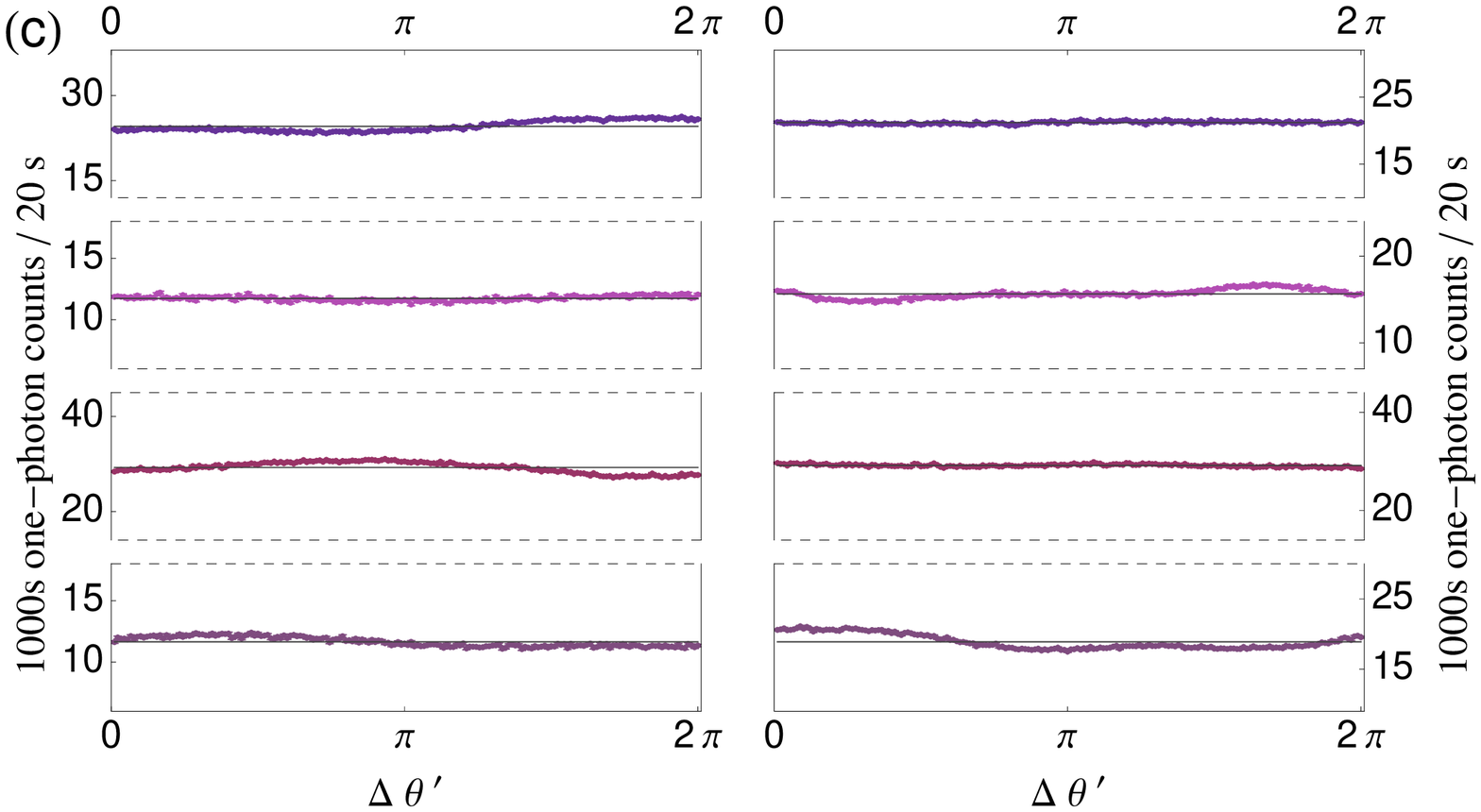}}
\vspace{-10 pt}
\caption{
Photonic quantum interference and one-photon response to the change in experimental geometric phase $\theta^{\prime}$.
(a) Raw data for all six coincidence outputs showing quantum interference as a function of $\Delta\theta^{\prime}$.
(b) The four $\Delta\theta^{\prime}$ sensitive quantum signals for a $2 \pi$ range of the geometric phase, normalised.
(c) Raw data for all eight output signals for both individual one-photon inputs.}
\label{fgFringes}
\vspace{-10 pt}
\end{figure*}

Devices implementing discrete instances of transformations similar to $H_{4}$ have been constructed in bulk optics for a small number of \emph{dynamical} phase settings \cite{ma-apb-60-S111} and in multimode waveguides \cite{pe-natcom-2-224} where the phase is permanently fixed at a single value.   Here, we encode in position (rail) and polarisation, using a pair of parallelised Jamin-Lebedeff interferometers in a calcite beam displacer architecture, as shown in Fig.~\ref{fgCirc}(b), to construct a device equivalent to $H_{4}$ up to trivial phases on input and output ports, and relabelling
\footnote
{Labelling a horizontally (vertically) polarised photon with $H(V)$, and a photon in the middle (bottom) rail with $m(b)$ the computational basis on inputs follows as
$
\ket{0}\equiv\ket{H,m},
\ket{1}\equiv\ket{V,b},
\ket{2}\equiv\ket{V,m},
\ket{3}\equiv\ket{H,b}
$.
Output labelling is indicated by detector subscripts.}.
Expanding to six modes through the $\theta^{\prime}$ section, we use an arrangement of waveplates to implement the phase geometrically, allowing continuous transition between all phase values.  The action of the beamsplitters in Fig.~\ref{fgCirc}(a) is equivalently implemented on polarisation by half waveplates ($A$) with optic axes set to $22.5^{\circ}$; the swap is facilitated with beam displacers ($B$) that force horizontally polarised light to \emph{walk off} at an angle, while vertically polarised light continues undeviated; polarisation beamsplitters ($S$) convert polarisation to position for detection.

The $\theta^{\prime}$ section in Fig.~\ref{fgCirc}(b) comprises quarter waveplates with optic axes fixed at $45^{\circ}$ ($Q$), or at $-45^{\circ}$ ($Q^{\dagger}$), and half waveplates ($P$) free to rotate but with optic axes locked together at the same angle $\alpha$.  Overall, this implements a polarisation flip on the top rail ($t$), the identity on the middle rail ($m$)
\footnote{The effective operation on the top and middle rails can easily be seen by noting that the two inner most quarter waveplates are mutually conjugate and, together, implement the identity.  Outside of these, the two half waveplates at the same angle also implement the identity, since a half waveplate gives an Hermitian operation.  Then the outer most quarter waveplates implement the polarisation flip (and fixed phase) on the top rail, and the identity on the middle rail.},
while the bottom rail ($b$) experiences a polarisation flip \emph{and} a vGP as a function of $\alpha$.  Figure~\ref{fgCirc}(c) shows the polarisation-sphere
\footnote{
Similar to the Bloch sphere or Poincar\'{e} sphere, canonical labelling on the sphere of light polarisation follows
$H(V)\equiv$ horizontal(vertical),
$D(A)\equiv$ diagonal(anti-diagonal),
$R(L)\equiv$ right(left) circular.
}.
Photons in rails $t$ and $m$, retrace their polarisation steps from the halfway point; while the total path length traversed on the polarisation-sphere for light in rail $b$ is equal to that of rails $t$ and $m$, the crucial difference in the actual route traversed leads to the experimental vGP, $\theta^{\prime}(\alpha)$.

The polarisation vector in rail $b$, of light travelling the $\theta^{\prime}$ section of Fig.~\ref{fgCirc}(b), makes a traversal of the sphere of Fig.~\ref{fgCirc}(c) that includes a closed cycle and an open path.  The full traversal is
VRP($\alpha$)LHRP($\alpha$)LH,
with P($\alpha$) indicating a point on the sphere determined by the variable angle of the half waveplates.  The respective $SU(2)$ unitaries for the polarisation vectors of the three rails are
\begin{align*}
U_{t}=i\begin{pmatrix} 0 & 1 \\ 1 & 0 \end{pmatrix},\;\;\;
U_{m}=\begin{pmatrix} 1 & 0 \\ 0 & 1 \end{pmatrix},\;\;\;
U_{b}=i \begin{pmatrix} 0 & e^{i 4 \alpha} \\ e^{-i 4 \alpha} & 0 \end{pmatrix},
\end{align*}
where $\ket{H}\equiv\ket{0}$ and $\ket{V}\equiv\ket{1}$ in each subspace.  In this construction, a $1^{\circ}$ change in the synchronised optic axes of both rotatable half waveplates results in a $4^{\circ}$ shift in the experimental geometric phase: $4\Delta\alpha=\Delta\theta^{\prime}$.  Light in rails $t$ and $b$ receive a polarisation flip during the full traversal so that the initial and final polarisation vectors of the respective rails are orthogonal.  Interestingly, a geometric phase is accrued in this situation \cite{ma-prl-85-3067} and can be identified as the fixed $i$ factor on these unitaries.

To observe the correlations predicted in Eq.~\eqref{eqProbs}, pairs of photons were injected into the network as shown in Fig.~\ref{fgCirc}(b): a horizontally polarised photon into rail $m$, and a vertically polarised photon into rail $b$, corresponding to computational states $\ket{0}$ and $\ket{1}$ respectively.  The $P$ waveplates were rotated almost two full revolutions, scanning a near $8\times 2\pi$ range for $\theta^{\prime}$.  To confirm the insensitivity of one-photon statistics to $\Delta\theta^{\prime}$, one photon was injected into the network, for each of the $\ket{0}$ and $\ket{1}$ inputs, with another photon sent directly to a detector, as a herald.  Photons were generated in a spontaneous parametric down conversion source \footnote{A $2$ mm thick BIBO crystal cut for Type $1$ phase matching was pumped with a $404$ $nm$ wavelength continuous wave laser, focused to a $40$ $\mu m$ waist.  Photons were collected, after passing through $2$ $nm$ interference filers, into polarisation maintaining optical fibres, and injected into the experimental network.}.

Experimental data shown in Fig.~\ref{fgFringes} strongly support predictions from Eq.~\eqref{eqProbs}.  Figure~\ref{fgFringes}(a) displays raw data for all six coincidence counts, with the expected four high visibility quantum interference fringes --- the difference in amplitudes is due to different coupling efficiencies.  In contrast, the two signals from detectors $D_{0}$ \& $D_{2}$ on rail $t$ (upper pane in Fig.~\ref{fgFringes}(a)), and $D_{1}$ \& $D_{3}$ on rail $m$ (lower pane in Fig.~\ref{fgFringes}(a)) show the predicted continuous and near total destructive quantum interference, with negligible response to $\Delta\theta^{\prime}$; the greatest phase response of these two signals was $D_{1}$ \& $D_{3}$ with a $0.9\pm0.1\%$ amplitude
\footnote{This amplitude was calculated by taking the difference between the visibilities in the HOM dips for the $\theta^{\prime}=0$ and $\theta^{\prime}=\pi$ cases.}.

Coincidence counts for the four $\Delta\theta^{\prime}$ quantum response curves were re-taken over the $2 \pi$ range shown in Fig.~\ref{fgFringes}(b) and normalised for the measured coupling efficiencies.  Following the notation from Eq.~\eqref{eqProbs},
the upper pane in Fig.~\ref{fgFringes}(b) shows $P(1 \mid 0)$ and $P(1 \mid 2)$ with a respective minimum and maximum at $\pi$ radians;
the lower pane in Fig.~\ref{fgFringes}(b) shows $P(3 \mid 2)$ and $P(3 \mid 0)$ with a respective minimum and maximum at $\pi$ radians.
The average visibility of these fringes is $94.0\pm0.2\%$ given from the fit that is shown as a solid black line.  The small flat data in Fig.~\ref{fgFringes}(b) are respective \emph{accidentals} signals, which are taken into account for the visibility calculation 
\footnote{\emph{Accidental} coincidence counts arise when the spontaneous parametric down conversion process (with a continuous wave pump) produces two photon pairs within the time resolution of the coincidence counter.  This rate can be directly measured by introducing a delay to omit all \emph{real} coincidences.}.
The raw data in Figure~\ref{fgFringes}(c) confirm the insensitivity of one-photon statistics to $\Delta\theta^{\prime}$.  The four panes on the left of Fig.~\ref{fgFringes}(c) show plots taken with a photon input in state $\ket{0}$ with the top plot showing data for detector $D_{0}$, the next plot below showing data for $D_{1}$ and so on; similarly the plots on the right of Fig.~\ref{fgFringes}(c) are taken for input state $\ket{1}$.  In each case the average count is shown as the solid line and the phase insensitivity is quantified by the relative standard error (RSE) from this line.  The average RSE from all eight plots is $3\%$; the plot with the greatest RSE at $6\%$ is from detector $D_{3}$ with input state $\ket{1}$.

The polarisation operations in the $\theta^{\prime}$ section of Fig.~\ref{fgCirc}(b) give zero dynamical phase contributions if the polarisation vectors are parallel transported at all times.  The condition for parallel transport, for a certain state vector $\ket{\psi(t)}$, is $\braket{\psi(t)}{\dt{\psi}(t)}=0$, which implies that, at an infinitesimal time step later, the sate $\ket{\psi(t + \delta t)}$ remains \emph{in phase} with $\ket{\psi(t)}$.  This condition is met for individual single photon states and for the composite two photon state superposed across all three rails of the $\theta^{\prime}$ section in Fig.~\ref{fgCirc}(b).  The lengths of rails $t$, $m$, and $b$ are not matched on the wavelength-scale of the photons so that unknown dynamical phase differences occur between these rails, which are fixed due to the intrinsic stability of the experimental architecture.  Therefore, the only variable phase is that on rail $b$ and is a function of the common adjustable angle of the two $P$ half waveplates.

We have observed photonic quantum interference fringes that are a function of a variable geometric phase.  This direct observation of the Berry phase in the quantum signal is possible because the optical complex Hadamard network contains no closed loops, so does not support single-photon interferometry.  All active elements of the state's Hilbert space make equal path length traversals on the polarisation sphere, but only those elements of the two-photon state that traverse via a particular route, lead to a variable geometric phase.  This route comprises a full cycle and an open path on the polarisation sphere, with mutually orthogonal start and end points.  We have therefore simultaneously experimentally tested several important and distinct aspects of the geometric phase in conjunction with making the central quantum-interference observation of the Berry phase.

The experimental circuit reported here is a small scale example of a model of quantum computation driven by photonic quantum interference \cite{pe-sci-329-1500, aaronsonQIP2010}, with a holonomic component.  Generalisations of holonomies have previously been widely investigated for robust gate operation in qubit based universal quantum computation, as discussed, and theoretically examined in the specific case of photonic qubits \cite{ma-pra-67-022316}.  Aside from imperfect unitary operation, the other major source of error in current linear optical experiments is typically photon loss, however, evidence suggests that loss in these models can be compensated through the injection of higher numbers of photons \cite{ro-arXiv1111.2426}. Furthermore, holonomies may reduce the exposure of photons to mode mismatch, leading to less demand on filtering, thus reducing loss.

Scaling up examples of holonomic multi-photon-interference-driven computational models in waveguides \cite{po-sci-320-646} where dynamical logic gates have been shown to work with high fidelity \cite{la-apl-97-211109} is an appealing prospect.  Any unitary transformation on modes can be implemented with a network of Mach Zehnder interferometers \cite{re-prl-73-58} which have been realised in waveguides with variable thermo-optic \cite{matthews-2008} and electro-optic \cite{bo-prl-108-053601} phase shifts, and integrated into a partially reconfigurable on-chip logic gate \cite{sh-nphot-6-45}. An interesting line of research is to consider the class of holonomic operations available given a large-scale fully reconfigurable optical unitary network, and the extent to which the global properties are invariant to imperfect splitting ratios in directional couplers and small random fluctuations from voltage controlled phase shifters, which act locally.

\begin{acknowledgments}
We thank S. Bartlett, N. Brunner, F. Flicker, M. Lobino, J. Matthews, K. Poulios, and P. Shadbolt 
for helpful discussions.  This work was supported by EPSRC, ERC, PHORBITECH and NSQI. J.L.O'B. acknowledges a Royal Society Wolfson Merit Award. 
\end{acknowledgments}


\begin{thebibliography}{49}
\expandafter\ifx\csname natexlab\endcsname\relax\def\natexlab#1{#1}\fi
\expandafter\ifx\csname bibnamefont\endcsname\relax
  \def\bibnamefont#1{#1}\fi
\expandafter\ifx\csname bibfnamefont\endcsname\relax
  \def\bibfnamefont#1{#1}\fi
\expandafter\ifx\csname citenamefont\endcsname\relax
  \def\citenamefont#1{#1}\fi
\expandafter\ifx\csname url\endcsname\relax
  \def\url#1{\texttt{#1}}\fi
\expandafter\ifx\csname urlprefix\endcsname\relax\def\urlprefix{URL }\fi
\providecommand{\bibinfo}[2]{#2}
\providecommand{\eprint}[2][]{\url{#2}}

\bibitem[{\citenamefont{Berry}(1983)}]{Berry}
\bibinfo{author}{\bibfnamefont{M.~V.} \bibnamefont{Berry}},
  \bibinfo{journal}{Proc. R. Soc. Lond. A} \textbf{\bibinfo{volume}{392}},
  \bibinfo{pages}{45} (\bibinfo{year}{1983}).
  
  \bibitem[{\citenamefont{Hong et~al.}(1987)\citenamefont{Hong, Ou, and
  Mandel}}]{ho-prl-59-2044}
\bibinfo{author}{\bibfnamefont{C.~K.} \bibnamefont{Hong}},
  \bibinfo{author}{\bibfnamefont{Z.~Y.} \bibnamefont{Ou}}, \bibnamefont{and}
  \bibinfo{author}{\bibfnamefont{L.}~\bibnamefont{Mandel}},
  \bibinfo{journal}{Phys. Rev. Lett.} \textbf{\bibinfo{volume}{59}},
  \bibinfo{pages}{2044} (\bibinfo{year}{1987}).
  
  \bibitem[{\citenamefont{Bromberg et~al.}(2009)\citenamefont{Bromberg, Lahini, Morandotti, and
  Silberberg}}]{br-prl-102-253904}
\bibinfo{author}{\bibfnamefont{Y.} \bibnamefont{Bromberg}},
  \bibinfo{author}{\bibfnamefont{Y.} \bibnamefont{Lahini}},
  \bibinfo{author}{\bibfnamefont{R.} \bibnamefont{Morandotti}}, \bibnamefont{and}
  \bibinfo{author}{\bibfnamefont{Y.}~\bibnamefont{Silberberg}},
  \bibinfo{journal}{Phys. Rev. Lett.} \textbf{\bibinfo{volume}{102}},
  \bibinfo{pages}{253904} (\bibinfo{year}{2009}).

\bibitem[{\citenamefont{Peruzzo et~al.}(2010)\citenamefont{Peruzzo, Lobino,
  Matthews, Matsuda, Politi, Poulios, Zhou, Lahini, Ismail, Wörhoff
  et~al.}}]{pe-sci-329-1500}
\bibinfo{author}{\bibfnamefont{A.}~\bibnamefont{Peruzzo}},
  \bibinfo{author}{\bibfnamefont{M.}~\bibnamefont{Lobino}},
  \bibinfo{author}{\bibfnamefont{J.~C.~F.} \bibnamefont{Matthews}},
  \bibinfo{author}{\bibfnamefont{N.}~\bibnamefont{Matsuda}},
  \bibinfo{author}{\bibfnamefont{A.}~\bibnamefont{Politi}},
  \bibinfo{author}{\bibfnamefont{K.}~\bibnamefont{Poulios}},
  \bibinfo{author}{\bibfnamefont{X.-Q.} \bibnamefont{Zhou}},
  \bibinfo{author}{\bibfnamefont{Y.}~\bibnamefont{Lahini}},
  \bibinfo{author}{\bibfnamefont{N.}~\bibnamefont{Ismail}},
  \bibinfo{author}{\bibfnamefont{K.}~\bibnamefont{Wörhoff}},
  \bibnamefont{et~al.}, \bibinfo{journal}{Science}
  \textbf{\bibinfo{volume}{329}}, \bibinfo{pages}{1500} (\bibinfo{year}{2010}).

\bibitem[{\citenamefont{Valiant}(1979)}]{va-tcs-8-189}
\bibinfo{author}{\bibfnamefont{L.~G.} \bibnamefont{Valiant}},
  \bibinfo{journal}{Theor. Comp. Sci.} \textbf{\bibinfo{volume}{8}},
  \bibinfo{pages}{189 } (\bibinfo{year}{1979}).

\bibitem[{\citenamefont{Scheel}(2004)}]{sc-arxiv-2004}
\bibinfo{author}{\bibfnamefont{S.}~\bibnamefont{Scheel}},
  \bibinfo{journal}{arXiv:quant-ph/0406127v1}  (\bibinfo{year}{2004}).

\bibitem[{\citenamefont{Aaronson and Arkhipov}(2010)}]{aaronsonQIP2010}
\bibinfo{author}{\bibfnamefont{S.}~\bibnamefont{Aaronson}} \bibnamefont{and}
  \bibinfo{author}{\bibfnamefont{A.}~\bibnamefont{Arkhipov}}, in
  \emph{\bibinfo{booktitle}{QIP 2010 - 13th Workshop on Quantum Information
  Processing}} (\bibinfo{year}{2010}).

\bibitem[{\citenamefont{Zanardi and Rasetti}(1999)}]{holonomic_proposal}
\bibinfo{author}{\bibfnamefont{P.}~\bibnamefont{Zanardi}} \bibnamefont{and}
  \bibinfo{author}{\bibfnamefont{M.}~\bibnamefont{Rasetti}},
  \bibinfo{journal}{Phys. Lett. A} \textbf{\bibinfo{volume}{264}},
  \bibinfo{pages}{94} (\bibinfo{year}{1999}).

\bibitem[{\citenamefont{Pachos et~al.}(1999)\citenamefont{Pachos, Zanardi, and
  Rasetti}}]{pa-pra-61-010305}
\bibinfo{author}{\bibfnamefont{J.}~\bibnamefont{Pachos}},
  \bibinfo{author}{\bibfnamefont{P.}~\bibnamefont{Zanardi}}, \bibnamefont{and}
  \bibinfo{author}{\bibfnamefont{M.}~\bibnamefont{Rasetti}},
  \bibinfo{journal}{Phys. Rev. A} \textbf{\bibinfo{volume}{61}},
  \bibinfo{pages}{010305} (\bibinfo{year}{1999}).

\bibitem[{\citenamefont{Jones et~al.}(2000)\citenamefont{Jones, Vedral, Ekert,
  and Castagnoli}}]{geometric_NMR}
\bibinfo{author}{\bibfnamefont{J.~A.} \bibnamefont{Jones}},
  \bibinfo{author}{\bibfnamefont{V.}~\bibnamefont{Vedral}},
  \bibinfo{author}{\bibfnamefont{A.}~\bibnamefont{Ekert}}, \bibnamefont{and}
  \bibinfo{author}{\bibfnamefont{G.}~\bibnamefont{Castagnoli}},
  \bibinfo{journal}{Nature} \textbf{\bibinfo{volume}{403}},
  \bibinfo{pages}{869} (\bibinfo{year}{2000}).

\bibitem[{\citenamefont{Duan et~al.}(2001)\citenamefont{Duan, Cirac, and
  Zoller}}]{dcz01}
\bibinfo{author}{\bibfnamefont{L.-M.} \bibnamefont{Duan}},
  \bibinfo{author}{\bibfnamefont{J.~I.} \bibnamefont{Cirac}}, \bibnamefont{and}
  \bibinfo{author}{\bibfnamefont{P.}~\bibnamefont{Zoller}},
  \bibinfo{journal}{Science} \textbf{\bibinfo{volume}{292}},
  \bibinfo{pages}{1695} (\bibinfo{year}{2001}).

\bibitem[{\citenamefont{Carollo et~al.}(2003)\citenamefont{Carollo,
  Fuentes-Guridi, Fran\c{c}a~Santos, and Vedral}}]{ca-prl-90-160402}
\bibinfo{author}{\bibfnamefont{A.}~\bibnamefont{Carollo}},
  \bibinfo{author}{\bibfnamefont{I.}~\bibnamefont{Fuentes-Guridi}},
  \bibinfo{author}{\bibfnamefont{M.}~\bibnamefont{Fran\c{c}a~Santos}},
  \bibnamefont{and} \bibinfo{author}{\bibfnamefont{V.}~\bibnamefont{Vedral}},
  \bibinfo{journal}{Phys. Rev. Lett.} \textbf{\bibinfo{volume}{90}},
  \bibinfo{pages}{160402} (\bibinfo{year}{2003}).

\bibitem[{\citenamefont{De~Chiara and Palma}(2003)}]{de-prl-91-090404}
\bibinfo{author}{\bibfnamefont{G.}~\bibnamefont{De~Chiara}} \bibnamefont{and}
  \bibinfo{author}{\bibfnamefont{G.~M.} \bibnamefont{Palma}},
  \bibinfo{journal}{Phys. Rev. Lett.} \textbf{\bibinfo{volume}{91}},
  \bibinfo{pages}{090404} (\bibinfo{year}{2003}).

\bibitem[{\citenamefont{Carollo et~al.}(2004)\citenamefont{Carollo,
  Fuentes-Guridi, Fran\c{c}a~Santos, and Vedral}}]{ca-prl-92-020402}
\bibinfo{author}{\bibfnamefont{A.}~\bibnamefont{Carollo}},
  \bibinfo{author}{\bibfnamefont{I.}~\bibnamefont{Fuentes-Guridi}},
  \bibinfo{author}{\bibfnamefont{M.}~\bibnamefont{Fran\c{c}a~Santos}},
  \bibnamefont{and} \bibinfo{author}{\bibfnamefont{V.}~\bibnamefont{Vedral}},
  \bibinfo{journal}{Phys. Rev. Lett.} \textbf{\bibinfo{volume}{92}},
  \bibinfo{pages}{020402} (\bibinfo{year}{2004}).

\bibitem[{\citenamefont{Solinas et~al.}(2004)\citenamefont{Solinas, Zanardi,
  and Zangh\`\i}}]{so-pra-70-042316}
\bibinfo{author}{\bibfnamefont{P.}~\bibnamefont{Solinas}},
  \bibinfo{author}{\bibfnamefont{P.}~\bibnamefont{Zanardi}}, \bibnamefont{and}
  \bibinfo{author}{\bibfnamefont{N.}~\bibnamefont{Zangh\`\i}},
  \bibinfo{journal}{Phys. Rev. A} \textbf{\bibinfo{volume}{70}},
  \bibinfo{pages}{042316} (\bibinfo{year}{2004}).

\bibitem[{\citenamefont{Fuentes-Guridi
  et~al.}(2005)\citenamefont{Fuentes-Guridi, Girelli, and
  Livine}}]{fu-prl-94-020503}
\bibinfo{author}{\bibfnamefont{I.}~\bibnamefont{Fuentes-Guridi}},
  \bibinfo{author}{\bibfnamefont{F.}~\bibnamefont{Girelli}}, \bibnamefont{and}
  \bibinfo{author}{\bibfnamefont{E.}~\bibnamefont{Livine}},
  \bibinfo{journal}{Phys. Rev. Lett.} \textbf{\bibinfo{volume}{94}},
  \bibinfo{pages}{020503} (\bibinfo{year}{2005}).

\bibitem[{\citenamefont{Filipp et~al.}(2009)\citenamefont{Filipp, Klepp,
  Hasegawa, Plonka-Spehr, Schmidt, Geltenbort, and Rauch}}]{fi-prl-102-030404}
\bibinfo{author}{\bibfnamefont{S.}~\bibnamefont{Filipp}},
  \bibinfo{author}{\bibfnamefont{J.}~\bibnamefont{Klepp}},
  \bibinfo{author}{\bibfnamefont{Y.}~\bibnamefont{Hasegawa}},
  \bibinfo{author}{\bibfnamefont{C.}~\bibnamefont{Plonka-Spehr}},
  \bibinfo{author}{\bibfnamefont{U.}~\bibnamefont{Schmidt}},
  \bibinfo{author}{\bibfnamefont{P.}~\bibnamefont{Geltenbort}},
  \bibnamefont{and} \bibinfo{author}{\bibfnamefont{H.}~\bibnamefont{Rauch}},
  \bibinfo{journal}{Phys. Rev. Lett.} \textbf{\bibinfo{volume}{102}},
  \bibinfo{pages}{030404} (\bibinfo{year}{2009}).

\bibitem[{\citenamefont{Pancharatnam}(1956)}]{pancharatnam1956}
\bibinfo{author}{\bibfnamefont{S.}~\bibnamefont{Pancharatnam}},
  \bibinfo{journal}{Proc. Ind. Acad. Sci. A} \textbf{\bibinfo{volume}{44}},
  \bibinfo{pages}{247 } (\bibinfo{year}{1956}).

\bibitem[{\citenamefont{Tomita and Chiao}(1986)}]{to-prl-57-937}
\bibinfo{author}{\bibfnamefont{A.}~\bibnamefont{Tomita}} \bibnamefont{and}
  \bibinfo{author}{\bibfnamefont{R.~Y.} \bibnamefont{Chiao}},
  \bibinfo{journal}{Phys. Rev. Lett.} \textbf{\bibinfo{volume}{57}},
  \bibinfo{pages}{937} (\bibinfo{year}{1986}).

\bibitem[{\citenamefont{Chyba et~al.}(1988)\citenamefont{Chyba, Wang, Mandel,
  and Simon}}]{ch-ol-13-562}
\bibinfo{author}{\bibfnamefont{T.~H.} \bibnamefont{Chyba}},
  \bibinfo{author}{\bibfnamefont{L.~J.} \bibnamefont{Wang}},
  \bibinfo{author}{\bibfnamefont{L.}~\bibnamefont{Mandel}}, \bibnamefont{and}
  \bibinfo{author}{\bibfnamefont{R.}~\bibnamefont{Simon}},
  \bibinfo{journal}{Optics Letters} \textbf{\bibinfo{volume}{13}},
  \bibinfo{pages}{562} (\bibinfo{year}{1988}).

\bibitem[{\citenamefont{Leek et~al.}(2007)\citenamefont{Leek, Fink, Blais,
  Bianchetti, Göppl, Gambetta, Schuster, Frunzio, Schoelkopf, and
  Wallraff}}]{le-sci-318-1889}
\bibinfo{author}{\bibfnamefont{P.~J.} \bibnamefont{Leek}},
  \bibinfo{author}{\bibfnamefont{J.~M.} \bibnamefont{Fink}},
  \bibinfo{author}{\bibfnamefont{A.}~\bibnamefont{Blais}},
  \bibinfo{author}{\bibfnamefont{R.}~\bibnamefont{Bianchetti}},
  \bibinfo{author}{\bibfnamefont{M.}~\bibnamefont{Göppl}},
  \bibinfo{author}{\bibfnamefont{J.~M.} \bibnamefont{Gambetta}},
  \bibinfo{author}{\bibfnamefont{D.~I.} \bibnamefont{Schuster}},
  \bibinfo{author}{\bibfnamefont{L.}~\bibnamefont{Frunzio}},
  \bibinfo{author}{\bibfnamefont{R.~J.} \bibnamefont{Schoelkopf}},
  \bibnamefont{and} \bibinfo{author}{\bibfnamefont{A.}~\bibnamefont{Wallraff}},
  \bibinfo{journal}{Science} \textbf{\bibinfo{volume}{318}},
  \bibinfo{pages}{1889} (\bibinfo{year}{2007}).

\bibitem[{\citenamefont{Wilczek and Zee}(1984)}]{wi-prl-52-2111}
\bibinfo{author}{\bibfnamefont{F.}~\bibnamefont{Wilczek}} \bibnamefont{and}
  \bibinfo{author}{\bibfnamefont{A.}~\bibnamefont{Zee}},
  \bibinfo{journal}{Phys. Rev. Lett.} \textbf{\bibinfo{volume}{52}},
  \bibinfo{pages}{2111} (\bibinfo{year}{1984}).

\bibitem[{\citenamefont{Aharonov and Anandan}(1987)}]{ah-prl-58-1593}
\bibinfo{author}{\bibfnamefont{Y.}~\bibnamefont{Aharonov}} \bibnamefont{and}
  \bibinfo{author}{\bibfnamefont{J.}~\bibnamefont{Anandan}},
  \bibinfo{journal}{Phys. Rev. Lett.} \textbf{\bibinfo{volume}{58}},
  \bibinfo{pages}{1593} (\bibinfo{year}{1987}).

\bibitem[{\citenamefont{Jordan}(1988)}]{jo-pra-38-1590}
\bibinfo{author}{\bibfnamefont{T.~F.} \bibnamefont{Jordan}},
  \bibinfo{journal}{Phys. Rev. A} \textbf{\bibinfo{volume}{38}},
  \bibinfo{pages}{1590} (\bibinfo{year}{1988}).

\bibitem[{\citenamefont{Wu and Li}(1988)}]{wu-prb-38-11907}
\bibinfo{author}{\bibfnamefont{Y.-S.} \bibnamefont{Wu}} \bibnamefont{and}
  \bibinfo{author}{\bibfnamefont{H.-Z.} \bibnamefont{Li}},
  \bibinfo{journal}{Phys. Rev. B} \textbf{\bibinfo{volume}{38}},
  \bibinfo{pages}{11907} (\bibinfo{year}{1988}).

\bibitem[{\citenamefont{Weinfurter and Badurek}(1990)}]{we-prl-64-1318}
\bibinfo{author}{\bibfnamefont{H.}~\bibnamefont{Weinfurter}} \bibnamefont{and}
  \bibinfo{author}{\bibfnamefont{G.}~\bibnamefont{Badurek}},
  \bibinfo{journal}{Phys. Rev. Lett.} \textbf{\bibinfo{volume}{64}},
  \bibinfo{pages}{1318} (\bibinfo{year}{1990}).

\bibitem[{\citenamefont{Samuel and Bhandari}(1988)}]{sa-prl-60-2339}
\bibinfo{author}{\bibfnamefont{J.}~\bibnamefont{Samuel}} \bibnamefont{and}
  \bibinfo{author}{\bibfnamefont{R.}~\bibnamefont{Bhandari}},
  \bibinfo{journal}{Phys. Rev. Lett.} \textbf{\bibinfo{volume}{60}},
  \bibinfo{pages}{2339} (\bibinfo{year}{1988}).

\bibitem[{\citenamefont{Manini and Pistolesi}(2000)}]{ma-prl-85-3067}
\bibinfo{author}{\bibfnamefont{N.}~\bibnamefont{Manini}} \bibnamefont{and}
  \bibinfo{author}{\bibfnamefont{F.}~\bibnamefont{Pistolesi}},
  \bibinfo{journal}{Phys. Rev. Lett.} \textbf{\bibinfo{volume}{85}},
  \bibinfo{pages}{3067} (\bibinfo{year}{2000}).

\bibitem[{\citenamefont{Lauber et~al.}(1994)\citenamefont{Lauber, Weidenhammer,
  and Dubbers}}]{la-prl-72-1004}
\bibinfo{author}{\bibfnamefont{H.-M.} \bibnamefont{Lauber}},
  \bibinfo{author}{\bibfnamefont{P.}~\bibnamefont{Weidenhammer}},
  \bibnamefont{and} \bibinfo{author}{\bibfnamefont{D.}~\bibnamefont{Dubbers}},
  \bibinfo{journal}{Phys. Rev. Lett.} \textbf{\bibinfo{volume}{72}},
  \bibinfo{pages}{1004} (\bibinfo{year}{1994}).

\bibitem[{\citenamefont{Hasegawa et~al.}(2001)\citenamefont{Hasegawa, Loidl,
  Baron, Badurek, and Rauch}}]{ha-prl-87-070401}
\bibinfo{author}{\bibfnamefont{Y.}~\bibnamefont{Hasegawa}},
  \bibinfo{author}{\bibfnamefont{R.}~\bibnamefont{Loidl}},
  \bibinfo{author}{\bibfnamefont{M.}~\bibnamefont{Baron}},
  \bibinfo{author}{\bibfnamefont{G.}~\bibnamefont{Badurek}}, \bibnamefont{and}
  \bibinfo{author}{\bibfnamefont{H.}~\bibnamefont{Rauch}},
  \bibinfo{journal}{Phys. Rev. Lett.} \textbf{\bibinfo{volume}{87}},
  \bibinfo{pages}{070401} (\bibinfo{year}{2001}).

\bibitem[{\citenamefont{Kwiat and Chiao}(1991)}]{kw-prl-66-588}
\bibinfo{author}{\bibfnamefont{P.~G.} \bibnamefont{Kwiat}} \bibnamefont{and}
  \bibinfo{author}{\bibfnamefont{R.~Y.} \bibnamefont{Chiao}},
  \bibinfo{journal}{Phys. Rev. Lett.} \textbf{\bibinfo{volume}{66}},
  \bibinfo{pages}{588} (\bibinfo{year}{1991}).

\bibitem[{\citenamefont{Strekalov and Shih}(1997)}]{st-pra-56-3129}
\bibinfo{author}{\bibfnamefont{D.~V.} \bibnamefont{Strekalov}}
  \bibnamefont{and} \bibinfo{author}{\bibfnamefont{Y.~H.} \bibnamefont{Shih}},
  \bibinfo{journal}{Phys. Rev. A} \textbf{\bibinfo{volume}{56}},
  \bibinfo{pages}{3129} (\bibinfo{year}{1997}).

\bibitem[{\citenamefont{Ericsson et~al.}(2005)\citenamefont{Ericsson, Achilles,
  Barreiro, Branning, Peters, and Kwiat}}]{er-prl-94-050401}
\bibinfo{author}{\bibfnamefont{M.}~\bibnamefont{Ericsson}},
  \bibinfo{author}{\bibfnamefont{D.}~\bibnamefont{Achilles}},
  \bibinfo{author}{\bibfnamefont{J.~T.} \bibnamefont{Barreiro}},
  \bibinfo{author}{\bibfnamefont{D.}~\bibnamefont{Branning}},
  \bibinfo{author}{\bibfnamefont{N.~A.} \bibnamefont{Peters}},
  \bibnamefont{and} \bibinfo{author}{\bibfnamefont{P.~G.} \bibnamefont{Kwiat}},
  \bibinfo{journal}{Phys. Rev. Lett.} \textbf{\bibinfo{volume}{94}},
  \bibinfo{pages}{050401} (\bibinfo{year}{2005}).

\bibitem[{\citenamefont{Klyshko}(1989)}]{kl-pla-140-19}
\bibinfo{author}{\bibfnamefont{D.}~\bibnamefont{Klyshko}},
  \bibinfo{journal}{Physics Letters A} \textbf{\bibinfo{volume}{140}},
  \bibinfo{pages}{19 } (\bibinfo{year}{1989}).

\bibitem[{\citenamefont{Brendel et~al.}(1995)\citenamefont{Brendel, Dultz, and
  Martienssen}}]{br-pra-52-2551}
\bibinfo{author}{\bibfnamefont{J.}~\bibnamefont{Brendel}},
  \bibinfo{author}{\bibfnamefont{W.}~\bibnamefont{Dultz}}, \bibnamefont{and}
  \bibinfo{author}{\bibfnamefont{W.}~\bibnamefont{Martienssen}},
  \bibinfo{journal}{Phys. Rev. A} \textbf{\bibinfo{volume}{52}},
  \bibinfo{pages}{2551} (\bibinfo{year}{1995}).

\bibitem[{\citenamefont{Kobayashi et~al.}(2011)\citenamefont{Kobayashi, Ikeda,
  Tamate, Nakanishi, and Kitano}}]{ko-pra-83-063808}
\bibinfo{author}{\bibfnamefont{H.}~\bibnamefont{Kobayashi}},
  \bibinfo{author}{\bibfnamefont{Y.}~\bibnamefont{Ikeda}},
  \bibinfo{author}{\bibfnamefont{S.}~\bibnamefont{Tamate}},
  \bibinfo{author}{\bibfnamefont{T.}~\bibnamefont{Nakanishi}},
  \bibnamefont{and} \bibinfo{author}{\bibfnamefont{M.}~\bibnamefont{Kitano}},
  \bibinfo{journal}{Phys. Rev. A} \textbf{\bibinfo{volume}{83}},
  \bibinfo{pages}{063808} (\bibinfo{year}{2011}).

\bibitem[{\citenamefont{Rarity et~al.}(1990)\citenamefont{Rarity, Tapster,
  Jakeman, Larchuk, Campos, Teich, and Saleh}}]{ra-prl-65-1348}
\bibinfo{author}{\bibfnamefont{J.~G.} \bibnamefont{Rarity}},
  \bibinfo{author}{\bibfnamefont{P.~R.} \bibnamefont{Tapster}},
  \bibinfo{author}{\bibfnamefont{E.}~\bibnamefont{Jakeman}},
  \bibinfo{author}{\bibfnamefont{T.}~\bibnamefont{Larchuk}},
  \bibinfo{author}{\bibfnamefont{R.~A.} \bibnamefont{Campos}},
  \bibinfo{author}{\bibfnamefont{M.~C.} \bibnamefont{Teich}}, \bibnamefont{and}
  \bibinfo{author}{\bibfnamefont{B.~E.~A.} \bibnamefont{Saleh}},
  \bibinfo{journal}{Phys. Rev. Lett.} \textbf{\bibinfo{volume}{65}},
  \bibinfo{pages}{1348} (\bibinfo{year}{1990}).

\bibitem[{\citenamefont{Tadej and \.{Z}yczkowski}(2006)}]{ta-osid-13-133}
\bibinfo{author}{\bibfnamefont{W.}~\bibnamefont{Tadej}} \bibnamefont{and}
  \bibinfo{author}{\bibfnamefont{K.}~\bibnamefont{\.{Z}yczkowski}},
  \bibinfo{journal}{OSID} \textbf{\bibinfo{volume}{13}}, \bibinfo{pages}{133}
  (\bibinfo{year}{2006}).

\bibitem[{\citenamefont{Durt et~al.}(2010)\citenamefont{Durt, Englert,
  Bengtsson, and \.{Z}yczkowski}}]{du-ijqi-8-535}
\bibinfo{author}{\bibfnamefont{T.}~\bibnamefont{Durt}},
  \bibinfo{author}{\bibfnamefont{B.-G.} \bibnamefont{Englert}},
  \bibinfo{author}{\bibfnamefont{I.}~\bibnamefont{Bengtsson}},
  \bibnamefont{and}
  \bibinfo{author}{\bibfnamefont{K.}~\bibnamefont{\.{Z}yczkowski}},
  \bibinfo{journal}{Int. J. Quant. Inf.} \textbf{\bibinfo{volume}{8}},
  \bibinfo{pages}{535} (\bibinfo{year}{2010}).

\bibitem[{\citenamefont{Mattle et~al.}(1995)\citenamefont{Mattle, Michler,
  Weinfurter, Zeilinger, and Zukowski}}]{ma-apb-60-S111}
\bibinfo{author}{\bibfnamefont{K.}~\bibnamefont{Mattle}},
  \bibinfo{author}{\bibfnamefont{M.}~\bibnamefont{Michler}},
  \bibinfo{author}{\bibfnamefont{H.}~\bibnamefont{Weinfurter}},
  \bibinfo{author}{\bibfnamefont{A.}~\bibnamefont{Zeilinger}},
  \bibnamefont{and} \bibinfo{author}{\bibfnamefont{M.}~\bibnamefont{Zukowski}},
  \bibinfo{journal}{Appl. Phys. B} \textbf{\bibinfo{volume}{60}},
  \bibinfo{pages}{S111} (\bibinfo{year}{1995}).

\bibitem[{\citenamefont{Peruzzo et~al.}(2011)\citenamefont{Peruzzo, Laing,
  Politi, Rudolph, and O'Brien}}]{pe-natcom-2-224}
\bibinfo{author}{\bibfnamefont{A.}~\bibnamefont{Peruzzo}},
  \bibinfo{author}{\bibfnamefont{A.}~\bibnamefont{Laing}},
  \bibinfo{author}{\bibfnamefont{A.}~\bibnamefont{Politi}},
  \bibinfo{author}{\bibfnamefont{T.}~\bibnamefont{Rudolph}}, \bibnamefont{and}
  \bibinfo{author}{\bibfnamefont{J.~L.} \bibnamefont{O'Brien}},
  \bibinfo{journal}{Nat Commun} \textbf{\bibinfo{volume}{2}},
  \bibinfo{pages}{224} (\bibinfo{year}{2011}).

\bibitem[{\citenamefont{Marzlin et~al.}(2003)\citenamefont{Marzlin, Bartlett,
  and Sanders}}]{ma-pra-67-022316}
\bibinfo{author}{\bibfnamefont{K.-P.} \bibnamefont{Marzlin}},
  \bibinfo{author}{\bibfnamefont{S.~D.} \bibnamefont{Bartlett}},
  \bibnamefont{and} \bibinfo{author}{\bibfnamefont{B.~C.}
  \bibnamefont{Sanders}}, \bibinfo{journal}{Phys. Rev. A}
  \textbf{\bibinfo{volume}{67}}, \bibinfo{pages}{022316}
  (\bibinfo{year}{2003}).

\bibitem[{\citenamefont{Rohde and Ralph}(2011)}]{ro-arXiv1111.2426}
\bibinfo{author}{\bibfnamefont{P.~P.} \bibnamefont{Rohde}} \bibnamefont{and}
  \bibinfo{author}{\bibfnamefont{T.~C.} \bibnamefont{Ralph}},
  \bibinfo{journal}{arXiv:1111.2426v1}  (\bibinfo{year}{2011}).

\bibitem[{\citenamefont{Politi et~al.}(2008)\citenamefont{Politi, Cryan,
  Rarity, Yu, and O'Brien}}]{po-sci-320-646}
\bibinfo{author}{\bibfnamefont{A.}~\bibnamefont{Politi}},
  \bibinfo{author}{\bibfnamefont{M.~J.} \bibnamefont{Cryan}},
  \bibinfo{author}{\bibfnamefont{J.~G.} \bibnamefont{Rarity}},
  \bibinfo{author}{\bibfnamefont{S.}~\bibnamefont{Yu}}, \bibnamefont{and}
  \bibinfo{author}{\bibfnamefont{J.~L.} \bibnamefont{O'Brien}},
  \bibinfo{journal}{Science} \textbf{\bibinfo{volume}{320}},
  \bibinfo{pages}{646} (\bibinfo{year}{2008}).

\bibitem[{\citenamefont{Laing et~al.}(2010)\citenamefont{Laing, Peruzzo,
  Politi, Verde, Halder, Ralph, Thompson, and O'Brien}}]{la-apl-97-211109}
\bibinfo{author}{\bibfnamefont{A.}~\bibnamefont{Laing}},
  \bibinfo{author}{\bibfnamefont{A.}~\bibnamefont{Peruzzo}},
  \bibinfo{author}{\bibfnamefont{A.}~\bibnamefont{Politi}},
  \bibinfo{author}{\bibfnamefont{M.~R.} \bibnamefont{Verde}},
  \bibinfo{author}{\bibfnamefont{M.}~\bibnamefont{Halder}},
  \bibinfo{author}{\bibfnamefont{T.~C.} \bibnamefont{Ralph}},
  \bibinfo{author}{\bibfnamefont{M.~G.} \bibnamefont{Thompson}},
  \bibnamefont{and} \bibinfo{author}{\bibfnamefont{J.~L.}
  \bibnamefont{O'Brien}}, \bibinfo{journal}{Applied Physics Letters}
  \textbf{\bibinfo{volume}{97}}, \bibinfo{eid}{211109}
  (pages~\bibinfo{numpages}{3}) (\bibinfo{year}{2010}).

\bibitem[{\citenamefont{Reck et~al.}(1994)\citenamefont{Reck, Zeilinger,
  Bernstein, and Bertani}}]{re-prl-73-58}
\bibinfo{author}{\bibfnamefont{M.}~\bibnamefont{Reck}},
  \bibinfo{author}{\bibfnamefont{A.}~\bibnamefont{Zeilinger}},
  \bibinfo{author}{\bibfnamefont{H.~J.} \bibnamefont{Bernstein}},
  \bibnamefont{and} \bibinfo{author}{\bibfnamefont{P.}~\bibnamefont{Bertani}},
  \bibinfo{journal}{Phys. Rev. Lett.} \textbf{\bibinfo{volume}{73}},
  \bibinfo{pages}{58} (\bibinfo{year}{1994}).

\bibitem[{\citenamefont{Matthews et~al.}(2009)\citenamefont{Matthews, Politi,
  Stefanov, and O'Brien}}]{matthews-2008}
\bibinfo{author}{\bibfnamefont{J.~C.~F.} \bibnamefont{Matthews}},
  \bibinfo{author}{\bibfnamefont{A.}~\bibnamefont{Politi}},
  \bibinfo{author}{\bibfnamefont{A.}~\bibnamefont{Stefanov}}, \bibnamefont{and}
  \bibinfo{author}{\bibfnamefont{J.~L.} \bibnamefont{O'Brien}},
  \bibinfo{journal}{Nature Photon.} \textbf{\bibinfo{volume}{3}},
  \bibinfo{pages}{346} (\bibinfo{year}{2009}).

\bibitem[{\citenamefont{Bonneau et~al.}(2012)\citenamefont{Bonneau, Lobino,
  Jiang, Natarajan, Tanner, Hadfield, Dorenbos, Zwiller, Thompson, and
  O'Brien}}]{bo-prl-108-053601}
\bibinfo{author}{\bibfnamefont{D.}~\bibnamefont{Bonneau}},
  \bibinfo{author}{\bibfnamefont{M.}~\bibnamefont{Lobino}},
  \bibinfo{author}{\bibfnamefont{P.}~\bibnamefont{Jiang}},
  \bibinfo{author}{\bibfnamefont{C.~M.} \bibnamefont{Natarajan}},
  \bibinfo{author}{\bibfnamefont{M.~G.} \bibnamefont{Tanner}},
  \bibinfo{author}{\bibfnamefont{R.~H.} \bibnamefont{Hadfield}},
  \bibinfo{author}{\bibfnamefont{S.~N.} \bibnamefont{Dorenbos}},
  \bibinfo{author}{\bibfnamefont{V.}~\bibnamefont{Zwiller}},
  \bibinfo{author}{\bibfnamefont{M.~G.} \bibnamefont{Thompson}},
  \bibnamefont{and} \bibinfo{author}{\bibfnamefont{J.~L.}
  \bibnamefont{O'Brien}}, \bibinfo{journal}{Phys. Rev. Lett.}
  \textbf{\bibinfo{volume}{108}}, \bibinfo{pages}{053601}
  (\bibinfo{year}{2012}).

\bibitem[{\citenamefont{Shadbolt et~al.}(2012)\citenamefont{Shadbolt, Verde,
  Peruzzo, Politi, Laing, Lobino, Matthews, Thompson, and
  O'Brien}}]{sh-nphot-6-45}
\bibinfo{author}{\bibfnamefont{P.~J.} \bibnamefont{Shadbolt}},
  \bibinfo{author}{\bibfnamefont{M.~R.} \bibnamefont{Verde}},
  \bibinfo{author}{\bibfnamefont{A.}~\bibnamefont{Peruzzo}},
  \bibinfo{author}{\bibfnamefont{A.}~\bibnamefont{Politi}},
  \bibinfo{author}{\bibfnamefont{A.}~\bibnamefont{Laing}},
  \bibinfo{author}{\bibfnamefont{M.}~\bibnamefont{Lobino}},
  \bibinfo{author}{\bibfnamefont{J.~C.~F.} \bibnamefont{Matthews}},
  \bibinfo{author}{\bibfnamefont{M.~G.} \bibnamefont{Thompson}},
  \bibnamefont{and} \bibinfo{author}{\bibfnamefont{J.~L.}
  \bibnamefont{O'Brien}}, \bibinfo{journal}{Nat Photon}
  \textbf{\bibinfo{volume}{6}}, \bibinfo{pages}{45} (\bibinfo{year}{2012}).

\end{thebibliography}
\end{document}